# COMPARATIVE ANALYSIS OF THE NUCLEOTIDE COMPOSITION BIASES IN EXONS AND INTRONS OF HUMAN GENES

Diana Duplij

*Institute of Molecular Biology and Genetics,
150 Academician Zabolotny Street, 03143 Kiev, Ukraine,
duplijd@gmail.com*

## Abstract

The nucleotide composition of human genes with a special emphasis on transcription-related strand asymmetries is analyzed. Such asymmetries may be associated with different mutational rates in two principal factors. The first one is transcription-coupled repair and the second one is the selective pressure related to optimization of the translation efficiency. The former factor affects both coding and noncoding regions of a gene, while the latter factor is applicable only to the coding regions. Compositional asymmetries calculated at the third position of a codon in coding (exons) and noncoding (introns, UTR, upstream and downstream) regions of human genes are compared. It is shown that the keto-skew (excess of the frequencies of G and T nucleotides over the frequencies of A and C nucleotides in the same strand) is most pronounced in intronic regions, less pronounced in coding regions, and has near zero values in untranscribed regions. The keto-skew correlates with the level of gene expression in germ-line cells in both introns and exons. We propose to use the results of our analysis to estimate the contribution of different evolutionary factors to the transcription-related compositional biases.

## 1. Background

A thorough analysis of base composition of genomic DNA started as early as DNA was discovered [1]. The Chargaff rule of base pairing (stating that fractions of adenines (A) and thymines (T) as well as fractions of guanines (G) and cytosines (C) are equal in genomic DNA) was first experimentally determined in native DNA and explained later when the DNA structure was described [2]. An extension of this rule (even approximately) to each strand of double-stranded DNA appears surprising [3,4] and the basis for this phenomenon is still debated.

The base composition parity rules were further studied and rigorously formulated [5,6]. It was shown that for the rule to be valid for each of the strands, the rates of direct and reverse single-nucleotide mutations should be equal for A and T, G and C, respectively. As soon as the sequencing of genomic DNA elaborated, significant deviations from this parity rule were reported [7,8]. Possible mechanisms responsible for the emergence of intrastrand asymmetries in the nucleotide composition were carefully reviewed by Lobry [9]. These mechanisms of arisen asymmetries which include those related to replication [10] and transcription [11].

The studying of nucleotide composition bias admits two main approaches [9]. The first approach anticipates using the phylogenetic reconstruction of nucleotide substitutions by analyzing homologous sequences from different species and tracing them to a common ancestor to determine the substitution rates of complementary nucleotides [12,13]. However, the data suitable for such investigations is rather scarce due to limited number of sequenced genomes. Another approach is based on the calculation of skew (biases) indices as a quantitative measure of the deviation of base composition in a single DNA strand from the parity rules [14,15,16]. The skew indices are computed as differences between single strand frequencies of certain nucleotides divided by the sum of the frequencies. For example, the GC-skew is measured as the value of ([G]-[C])/([C]+[G]) and AT-skew is ([A]-[T])/([A]+[T]), where brackets denote the absolute values of correspondent nucleotides. The skew method has proven to be useful in analyzing the base composition bias related to both DNA replication [10,17,18] and transcription processes [19,20,21,22].

Since the base composition bias originates from DNA mutations and it is controlled by selective pressure, one can expect that the degree of the bias differs for distinct functional regions of the genome. Indeed, it was shown that the base-composition skew, keto-skew in particular, is significantly different in exons and introns of human genes [23,24]. It has been also shown that the value of the skew in introns correlates with the transcription level of the corresponding genes measured in human oocytes. [25].

In this paper we present a comparative analysis of the relation between skews in both exons and introns and the transcription level. We demonstrate the apparent correlation between internal introns keto-skew with the expression level of human genes in testis germ cell.

## 2. Results and Discussion

The pronounced differences in base-pair composition skew for exons and introns of human genes were previously demonstrated [23,24]. For instance, the keto-skew value ($K_s$= ([G]+[T]-[A]-[C])/ ([G]+[T]+[A]+[C])) is negative in exons and positive in introns. The keto-skew is associated to the transcription-related mutation events which are among the most frequent ones in human genome [26]. Here we focus our current analysis on this kind of indices.

It is clear from Fig. 1 that the values of $K_s$ averaged over the internal exons differ significantly from those on introns.

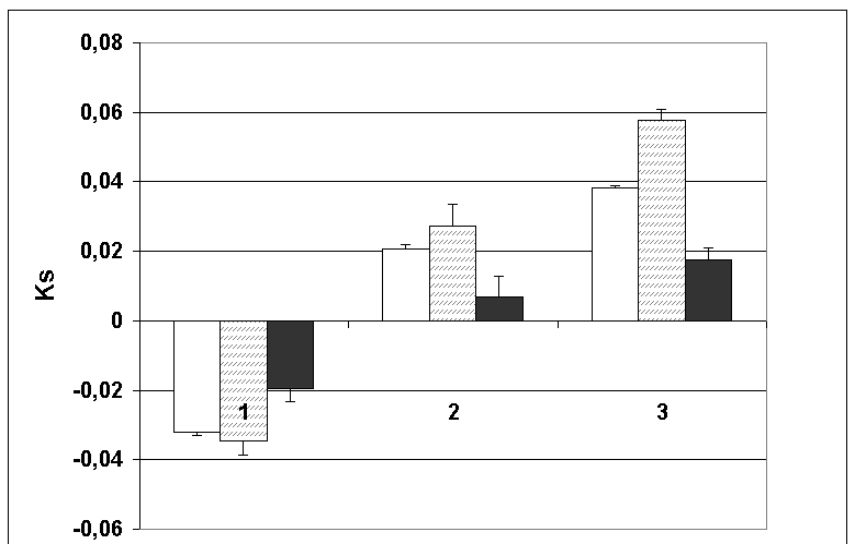

Fig.1. $K_s$ values in CDS third positions of codons and internal introns of different groups of genes.

On Fig. 1 each column represents averaged $K_s$ values over the following gene groups: white columns identify common group genes (6472), shaded columns identify housekeeping genes (266), and black columns identify tissue-specific genes (197) (see the section "Materials and methods"). The standard errors are also depicted. The functional triples of gene regions from each mentioned groups are enumerated by: 1. Internal exons of gene mRNA. 2. The third position of CDS codons. 3. Internal introns of gene mRNA.
At the same time, the exonic values of $K_s$ averaged only over the 3rd positions of codons are closer to the intronic values. This observation agrees with the fact that the $3^{rd}$ position in codons is normally under weaker selective pressure than the $1^{st}$ and $2^{nd}$ positions [27]. Our results suggest, however, that the selective pressure acting on the $3^{rd}$ position of codons is still higher than the pressure acting on introns (c.f. the absolute values of $K_s$ for introns and 3rd positions of codons in Fig.1).

It is also observed that the absolute values of $K_s$ are higher for housekeeping genes than they are for tissue-specific genes in both exons and introns. For instance, the mean $K_s$ in the $3^{rd}$ codon positions in housekeeping genes (0.0272±0.0063) is almost four times higher than the corresponding value in tissue-specific genes (0.0069±0.0059). Also, $K_s$ of introns in housekeeping genes (0.058±0.0029) is 3.4 times higher than the $K_s$ of tissue-specific introns (0.017±0.0036). The housekeeping genes are highly expressed in broad range of tissues and such behavior of keto-skew can reflect the dependence of this index on transcription level.

Next, we analyze the specific dependence of $K_s$ on the transcriptional activity at the level of individual genes. As shown in Fig. 2, the $K_s$ values correlate moderately with the gene expression levels.

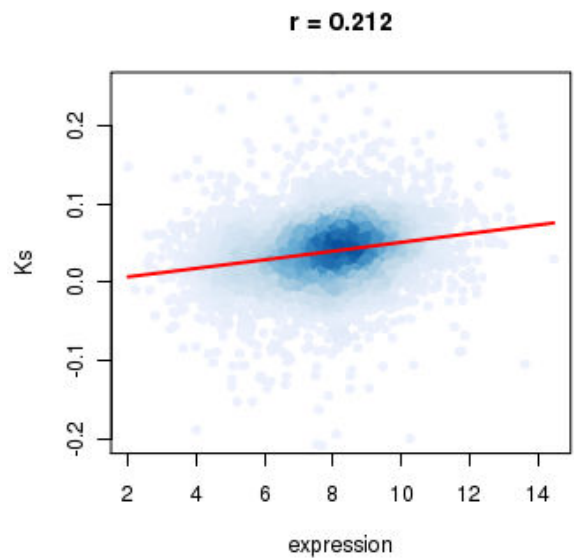

Fig.2. Correlation between internal introns $K_s$ values and expression's level of 6472 genes.

As compared to the corresponding values computed for all genes in our data set, the correlation is more pronounced for the subset of housekeeping genes and less pronounced for the subset of tissue-specific genes. In the group of genes expressed in hypothetical product the correlation is negative (Table 1). This dependence confirms our observation that $K_s$ value is higher on average in housekeeping genes than it is in tissue-specific genes (Fig. 1). Additionally the genes associated with malignant melanoma were taken for account as an example of the intensively expressed genes [28]. The internal intron $K_s$ values obtained in the somatic cancer genes (0.0516 ±0.0014) and cancer germ line genes (0.0508±0.0025) are comparable, but these values do not exceed one in housekeeping introns (0.0579±0.0029).

Thus the nucleotide composition of human genes with different expression patterns has been analyzed. It the researched reveals that the intronic complementary nucleotide biases are substantially higher than exonic ones. The representative factor here is the keto-skew value, which reflects the specific bias between quantities of nucleotides C+A ↔ T+G. This value allows one to trace the relation between most frequently mutating nucleotides C ↔T and G ↔A. It is also discovered, that in loci with least selective pressure, the keto-skew is shifted toward the predominance of T and A. Likewise, as it was demonstrated previously in [24], the keto-skew values are close to zero in 5'-UTR of human genes.

Another observation is a moderate correlation between keto-skew values and the expression level. In view of that, an extensive transcription results in the accumulation of transitions C >T and G >A in introns. Probably this happens due the RNA polymerase ability to avoid stalls in non-coding DNA lesions [22]. On the other hand, the known nucleotide excision repair mechanisms aimed precisely at transcription-blocking lesions [21]. It was shown [28], that C>T/G>A transitions constitute the most commonest class of substitutions in melanoma. However, the effect of transcription-coupled repair account only for one-third of the deficit of mutations over protein-coding gene footprints [28]. The $K_s$ values computed for the housekeeping and cancer genes in our data set are very close. Probably some mechanisms constrain the skew values elevation and this needs the next thorough investigation.

We conclude that the keto-skew value allows one to improve our understanding of mechanisms of transcription-coupled spontaneous mutagenesis.

## 3. Material and Methods

The human genome (hg17) was downloaded from the National Center for Biotechnology Information (NCBI) archive (ftp://ftp.ncbi.nih.gov/genbank/genomes/H_sapiens). The female genome model which comprised all autosomes and the chromosome X was considered in the current study. The GenBank genes annotations were used to determine genes and exons boundaries. If several versions of CDS were present, only the most upstream ones were considered.

We used several filtering steps to obtain the non-redundant and high-confidence gene set. First, highly homologous lymphocyte and immunoglobulin genes were removed from the list. Then, all one-exon genes and genes with the length of any exon or intron less than 30 bp were filtered out to omit extreme values of skew indices before averaging. After filtering, the dataset comprised 10839 genes. For 6472 of those, the values of expression in testis germ cells were estimated from the published microarray data (GEO ID GSM18985 and GSM18986) [http://www.ncbi.nlm.nih.gov/geo/]. These genes were used as a final set for calculations of skew indices. The 10 bp fragments at the ends of introns and exons were discarded from the calculation of skew indices to avoid possible bias related to the presence of highly conserved sequences in these regions. The published lists of housekeeping and tissue-specific genes [29,30] were used in our analysis. Those genes which have in their product definition the word "hypothetical" were mark out. Gene group associated with malignant melanoma [28] were also separated. We used the supplementary table of [28] and filtered out 259 somatic cancer genes and 66 germ line cancer genes.

**Acknowledgments**

Author would like to thank M. Tolstorukov for principal inspiration and many useful discussions.

Table 1. The values of $K_s$ and correlation coefficients for different gene groups.

| Gene group | Correlation coefficient Ks value with expression level | Mean $K_s$ value of internal introns |
|---|---|---|
| Common group | 0.2120664 | 0.038204±0.000524 |
| Housekeeping | 0.2405525 | 0.057859±0.002891 |
| Tissue-specific | 0.1730654 | 0.017401±0.00356 |
| Hypothetical | -0.0616326 | 0.035907±0.000177 |
| Rest genes | 0.2196712 | 0.038816±0.00565 |